\documentclass[fleqn,usenatbib]{mnras}

\usepackage{newtxtext,newtxmath}
\DeclareRobustCommand{\VAN}[3]{#2}
\let\VANthebibliography\thebibliography
\def\thebibliography{\DeclareRobustCommand{\VAN}[3]{##3}\VANthebibliography}

\usepackage{graphicx}	% Including figure files
\usepackage{amsmath}	% Advanced maths commands
\def \aj {AJ}
\def \mnras {MNRAS}
\def \pasp {PASP}
\def \apj {ApJ}
\def \apjs {ApJS}
\def \apjl {ApJL}
\def \aap {A\&A}
\def \nat {Nature}
\def \araa {ARAA}

\def\lesssim{\mathrel{\hbox{\rlap{\hbox{\lower4pt\hbox{$\sim$}}}\hbox{$<$}}}}
\def\gtrsim{\mathrel{\hbox{\rlap{\hbox{\lower4pt\hbox{$\sim$}}}\hbox{$>$}}}}

\title[Polarization of AT~2018cow]{A flash of polarized optical light points to an aspherical ``cow''}
\author[Maund et al.]{Justyn R. Maund$^{1},$\thanks{Email: j.maund@sheffield.ac.uk} Peter A. H\"{o}flich,$^{2}$ Iain A Steele,$^{3}$ Yi Yang,$^{4}$\thanks{Bengier-Winslow-Robertson Fellow} Klaas Wiersema, $^{5,6}$\newauthor Shiho Kobayashi,$^{3}$ Nuria Jordana-Mitjans, $^{7}$  Carole Mundell,$^{7}$ Andreja Gomboc,$^{8}$ \newauthor Cristiano Guidorzi,$^{9,10,11}$ and Robert J. Smith$^{3}$\\
$^{1}$Department of Physics and Astronomy, University of Sheffield, Hicks Building, Hounsfield Road, Sheffield, S3 7RH, UK\\
$^{2}$Department of Physics, Florida State University, Tallahassee, FL 32306, USA\\
$^{3}$ Astrophysics Research Institute, Liverpool John Moores University, 146 Brownlow Hill, Liverpool, L3 5RF, UK\\
$^{4}$ Department of Astronomy, University of California, Berkeley, CA 94720-3411, USA \\
$^{5}$ Physics Department, Lancaster University, Lancaster, LA1 4YB, UK\\
$^{6}$ Department of Physics, University of Warwick, Gibbet Hill Road, Coventry CV4 7AL, UK\\
$^{7}$ Department of Physics, University of Bath, Claverton Down, Bath BA2 7AY, UK\\
$^{8}$ Center for Astrophysics and Cosmology, University of Nova Gorica, Vipavska 13, 5000 Nova Gorica, Slovenia\\
$^{9}$ INFN – Sezione di Ferrara, Via Saragat 1, I-44122 Ferrara, Italy\\
$^{10}$ INAF – Osservatorio di Astrofisica e Scienza dello Spazio di Bologna, Via Piero Gobetti 101, I-40129 Bologna, Italy\\
$^{11}$ INAF – Osservatorio Astronomico di Cagliari, via della Scienza 5, I-09047 Selargius, Italy
}

\date{Accepted XXX. Received YYY; in original form ZZZ}

\pubyear{2022}

\begin{document}
\label{firstpage}
\pagerange{\pageref{firstpage}--\pageref{lastpage}}
\maketitle

\begin{abstract}
The astronomical transient AT2018cow is the closest example of the new class of luminous, fast blue optical transients (FBOTs).  Liverpool Telescope {\it RINGO3} observations of AT~2018cow are reported here, which constitute the earliest polarimetric observations of an FBOT.  At $5.7\,\mathrm{days}$ post-explosion, the optical emission of AT2018cow exhibited a chromatic polarization spike that reached $\sim 7\%$ at red wavelengths.  This is the highest intrinsic polarization recorded for a non-relativistic explosive transient, and is observed in multiple bands and at multiple epochs over the first night of observations, before rapidly declining.   The apparent wavelength dependence of the polarization may arise through depolarization or dilution of the polarized flux, due to conditions in AT~2018cow at early times.  A second ``bump" in the polarization is observed at blue wavelengths at $\sim 12\,\mathrm{days}$.  Such a high polarization requires an extremely aspherical geometry that is only apparent for a brief period ($<1$ day), such as shock breakout through an optically thick disk.  For a disk-like configuration, the ratio of the thickness to radial extent must be $\sim 10\%$.

\end{abstract}

\begin{keywords}
technique:polarimetric -- stars -- supernovae:individual:AT~2018cow
\end{keywords}

%%%%%%%%%%%%%%%%%%%%%%%%%%%%%%%%%%%%%%%%%%%%%%%%%%

%%%%%%%%%%%%%%%%% BODY OF PAPER %%%%%%%%%%%%%%%%%%

%%%%%%%%%%%%%%%%%%%%%%%%%%%%%%%%%%%%%%%%%%%%%%%%%%%%%%%%%%%%%%%%%%%%%%
% INTRODUCTION - INTRODUCTION - INTRODUCTION - INTRODUCTION
% INTRODUCTION - INTRODUCTION - INTRODUCTION - INTRODUCTION
% INTRODUCTION - INTRODUCTION - INTRODUCTION - INTRODUCTION
% INTRODUCTION - INTRODUCTION - INTRODUCTION - INTRODUCTION
%%%%%%%%%%%%%%%%%%%%%%%%%%%%%%%%%%%%%%%%%%%%%%%%%%%%%%%%%%%%%%%%%%%%%%

\section{Introduction}

%Introduce the FBOTs
The advent of deep, wide-field, high-cadence surveys in the last decade, such as the Zwicky Transient Facility \citep[ZTF;][]{2019pasp..131a8002b}, the Asteroid Terrestrial-Impact Last Alert System \citep[ATLAS;][]{2011pasp..123...58t}, the All-Sky Automated Survey for Supernovae \citep[ASAS-SN;][]{2014apj...788...48s}, and the Gravitational-wave Optical Transient Observer \citep[GOTO;][]{2020mnras.497..726g}, has yielded discoveries of fast evolving transients at cosmological distances that had been previously missed by lower cadence, shallower surveys.  The fast blue optical transients (FBOTs) are characterised by rapid evolution, with a rise to peak luminosity on timescales $\lesssim 10\,\mathrm{days}$ \citep[][]{2014apj...794...23d,2016apj...819...35a,2018mnras.481..894p} and high luminosities, briefly comparable to those of superluminous supernovae \citep{2019ara&a..57..305g}.  The rapid evolution of luminous FBOTs  cannot be explained solely through the decay of radioactive nuclides, in particular $\mathrm{^{56}Ni}$, as found for normal supernovae \citep{2014apj...794...23d,2018mnras.481..894p}.

%Details of AT2018cow as prototype

At the bright extreme, the prototype of the luminous FBOTs is AT~2018cow.  Although FBOTs are more usually found at high redshift \citep{2014apj...794...23d}, AT2018cow (also designated ATLAS18qqn) was discovered in the galaxy CGCG 137-068 at a distance of only $60\,\mathrm{Mpc}$ \citep{2018atel11727....1s}, making it the closest and most well-studied FBOT, being subject to multiwavelength scrutinity over the course of its rapid evolution \citep{2018apj...865l...3p, 2019apj...872...18m, 2019mnras.484.1031p}.  AT2018cow was observed to increase rapidly in brightness by $>5\,\mathrm{mags}$ in only 3.5 days and then decline rapidly \citep[$\Delta m_{15}(g)\sim 3\,\mathrm{mags}$; ][]{2018apj...865l...3p}.  For the first 20 days the spectrum was almost featureless, characterised by a blue continuum corresponding to a black body with temperature  $T \approx 50\,000\mathrm{K}$ \citep{2018ATel11740....1X, 2018atel11776....1p} similar to broad-lined Type Ic SNe \citep{2018atel11753....1i,2001apj...555..900p}.  After $\sim 20\,\mathrm{days}$, the spectrum of AT2018cow evolved and features identified as being due to hydrogen and helium began to emerge \citep{2018apj...865l...3p, 2019apj...872...18m, 2019mnras.484.1031p}.  With the identification of other AT~2018cow-like events, such as CSS161010 \citep{2020apj...895l..23c}, AT~2018lug \citep{2020apj...895...49h}, AT~2020mrf \citep{2022apj...934..104y} and AT~2020xnd \citep{2021mnras.508.5138p, 2022apj...926..112b}, such luminous transients are emerging as their own class of astrophysical phenomenon rather than being an extreme extension of the already established varieties of  supernovae.

A diverse range of scenarios have been proposed to explain the peculiar behaviour of AT~2018cow and similar FBOTs: the failed explosion of a supergiant star that forms a compact object, which accretes material from the progenitor \citep{2019apj...872...18m, 2019mnras.485l..83q}; an electron-capture SN resulting from a merger of white dwarfs \citep{2019mnras.487.5618l}; a tidal disruption event \citep{2019mnras.487.2505k,2019mnras.484.1031p}, although associated with an intermediate or stellar mass blackhole, due to both the location of AT~2018cow in its host galaxy \citep[see][]{2022arxiv221001144s} and the fast rise-time \citep{2019mnras.487.2505k,2019apj...871...73h}; or a jet originating from a merger of a neutron star and a massive star evolving in a common envelope \citep{2022raa....22e5010s}.  The later appearance of H and He lines led to comparisons with interacting Type Ibn SNe \citep{2019mnras.488.3772f}.  The near universal location of these events in star-forming, dwarf galaxies ($M_{\star} \sim 10^{7-9}M_{\odot}$) has led to the association of FBOTs with massive stars \citep[see ][ and references therein]{2022apj...934..104y}.

The development of a complete physical picture for FBOTs has been hampered by how rare they are, with the X-ray and radio bright FBOTs occuring at $<1\%$ of the local core-collapse SN rate \citep{2020apj...895l..23c,2022apj...926..112b,2020apj...894...27t}.  Despite the identification of four analogues to AT~2018cow (given above), the prototype remains the most comprehensively monitored.
Given the lack of evidence for the role of radioactive nickel in these events, an accreting central engine (either neutron star or black hole) has emerged as a leading requirement for AT~2018cow \citep{2019apj...872...18m}.  \citet{2022natas...6..249p} report the detection of a quasi-periodic oscillation in soft X-rays from AT~2018cow with a timescale of 4.4ms, indicating the presence of a neutron star or blackhole.  Asymmetries, or rather departures from spherical symmetry, have also been indirectly inferred from the observations due to: the presence of constrasting velocity regimes, with distinct polar ($v \sim 0.1c$) and equatorial ($\sim 6000\,\mathrm{km\,s^{-1}}$) flows \citep{2019apj...872...18m, 2019mnras.484.1031p, 2019mnras.488.3772f}; the late-time H and He emission lines appearing to be systematically redshifted by $3000\,\mathrm{km\,s^{-1}}$ \citep{2018apj...865l...3p, 2019mnras.484.1031p, 2019mnras.488.3772f}; and the steep decline of the luminosity at mm-wavelengths \citep{2019apj...871...73h}.

%Role of polarimetry for assessing the asymmetry
As has been demonstrated for SNe \citep{2008ara&a..46..433w}, tidal disruption events \citep{2022natas...6.1193l} and kilonovae \citep{2019natas...3...99b}, polarimetry has the power to probe the geometries of transient phenomena; in particular, through the polarization induced through Thomson scattering from free-electrons.  Polarimetry may represent, therefore, one of the key observational constraints with which to test the geometries required for  the different scenarios proposed for FBOTs and, in particular, AT~2018cow.

The first time-series of broad-band polarimetric observations of an FBOT, AT~2018cow, are presented here.  In Section \ref{sec:obs} we present the Liverpool Telescope {\it RINGO3} observations of AT~2018cow and we present the polarimetric measurements (and establish their significance) in Section \ref{sec:res}.  We discuss the implications of these observations in the context of other observations and proposed models for AT~2018cow in Section \ref{sec:discussion}.

%%%%%%%%%%%%%%%%%%%%%%%%%%%%%%%%%%%%%%%%%%%%%%%%%%%%%%%%%%%%%%%%%%%%%%
% OBSERVATIONS - OBSERVATIONS - OBSERVATIONS - OBSERVATIONS
% OBSERVATIONS - OBSERVATIONS - OBSERVATIONS - OBSERVATIONS
% OBSERVATIONS - OBSERVATIONS - OBSERVATIONS - OBSERVATIONS
% OBSERVATIONS - OBSERVATIONS - OBSERVATIONS - OBSERVATIONS
%%%%%%%%%%%%%%%%%%%%%%%%%%%%%%%%%%%%%%%%%%%%%%%%%%%%%%%%%%%%%%%%%%%%%%

\section{Observations}
\label{sec:obs}
Polarimetric observations of AT2018cow, using the Liverpool Telescope \citep{2004spie.5489..679s} {\it RINGO3} imaging polarimeter \citep{2012spie.8446e..2ja}, commenced on the night of 2018 June 20, or 5.6 days after the estimated explosion date \citep[2018 June 15 07:12:00UTC; ][]{2018apj...865l...3p}.  Observations continued until 2018 July 21, corresponding to 36.6 days post-explosion.  A log of the {\it RINGO3} observations of AT~2018cow is presented in Table \ref{tab:obs:pol}.

The {\it RINGO3} polarimeter consisted of a rapidly rotating wire grid polarizer, which selects the polarization component to be measured at any one time.  The beam of light was then depolarized using a Lyot prism and was then directed through two dichroic mirrors into 3 separate cameras covering 3 wavelength ranges: $b^{\ast}$: $4263 - 6495\mathrm{\mathring{A}}$, $g^{\ast}$ : $6456 - 7586\mathrm{\mathring{A}}$ and $r^{\ast}$ : $7601 - 8436\mathrm{\mathring{A}}$  \citep{2012spie.8446e..2ja,2020apj...892...97j}.  A complete sequence of measurements, with the polarizer at 8 positions, was completed every 2.3s simultaneously in all three wavelength channels.

\begin{table*}
\caption{Liverpool Telescope {\it RINGO3} observations of the polarization of AT~2018cow\label{tab:obs:pol}}
\begin{tabular}{lcccccc}
    \hline
             &.         &                   &                   & \multicolumn{3}{c}{Polarization$^{\ddagger}$} \\
    \cline{5-7}
    Date & MJD & Days since & Exposure  & $p(b^{\ast})$ & $p(g^{\ast})$ & $p(r^{\ast})$ \\
    (UTC) &     & explosion$^{\dagger}$ & Time (s) & ($\%$) & $(\%)$ & ($\%$) \\
    \hline
 2018-06-20 21:40 & 58289.90 & 5.60  & 195 & $1.16\pm 0.30$ & $3.16\pm 0.69$ & $3.93\pm 1.27$\\
 2018-06-20 22:24 & 58289.93 & 5.63  & 596 & $1.21\pm 0.18$ & $3.70\pm 0.44$ & $5.06\pm 0.81$\\
 2018-06-21 01:54 & 58290.08 & 5.78  & 197 & $2.20\pm 0.27$ & $4.81\pm 0.63$ & $6.89\pm 1.37$\\
 2018-06-21 21:45 & 58290.91 & 6.61  & 197 & $1.30\pm 0.43$ & $2.08\pm 0.98$ & $3.55\pm 1.88$\\
 2018-06-21 23:43 & 58290.99 & 6.69  & 298 & $0.32\pm 0.37$ & $0.74\pm 0.76$ & $0.88\pm 1.56$\\
 2018-06-22 01:54 & 58291.08 & 6.78  & 197 & $0.71\pm 0.37$ & $0.00\pm 0.89$ & $1.71\pm 1.81$\\
 2018-06-22 22:21 & 58291.93 & 7.63  & 197 & $0.94\pm 0.53$ & $0.58\pm 1.18$ & $0.00\pm 2.21$\\
 2018-06-23 02:28 & 58292.10 & 7.80  & 197 & $0.15\pm 0.47$ & $1.84\pm 1.18$ & $2.12\pm 2.14$\\
 2018-06-23 21:15 & 58292.89 & 8.59  & 197 & $0.28\pm 0.67$ & $0.14\pm 1.67$ & $0.05\pm 2.71$\\
 2018-06-23 23:13 & 58292.97 & 8.67  & 598 & $0.01\pm 0.42$ & $0.70\pm 0.89$ & $0.17\pm 1.75$\\
 2018-06-24 01:23 & 58293.06 & 8.76  & 197 & $1.05\pm 0.66$ & $2.37\pm 1.35$ & $0.00\pm 2.96$\\
 2018-06-24 21:58 & 58293.92 & 9.62  & 197 & $0.89\pm 0.81$ & $2.09\pm 1.79$ & $2.77\pm 3.04$\\
 2018-06-25 21:19 & 58294.89 & 10.59 & 197 & $1.43\pm 1.25$ & $1.27\pm 2.05$ & $0.65\pm 3.59$\\
 2018-06-26 01:33 & 58295.07 & 10.77 & 197 & $0.43\pm 1.20$ & $1.69\pm 2.20$ & $3.21\pm 3.95$\\
 2018-06-26 21:42 & 58295.90 & 11.60 & 197 & $0.00\pm 1.13$ & $1.88\pm 2.61$ & $0.00\pm 4.23$\\
 2018-06-27 01:58 & 58296.08 & 11.78 & 197 & $0.00\pm 1.47$ & $4.07\pm 2.33$ & $0.00\pm 4.39$\\
 2018-06-27 21:17 & 58296.89 & 12.59 & 197 & $0.00\pm 1.23$ & $1.30\pm 2.16$ & $0.00\pm 3.86$\\
 2018-06-28 01:26 & 58297.06 & 12.76 & 197 & $3.30\pm 1.36$ & $0.03\pm 2.14$ & $0.00\pm 3.21$\\
 2018-06-28 21:56 & 58297.91 & 13.61 & 596 & $1.91\pm 0.72$ & $0.00\pm 1.25$ & $0.37\pm 2.20$\\
 2018-06-28 23:06 & 58297.96 & 13.66 & 596 & $1.40\pm 0.67$ & $0.27\pm 1.23$ & $2.81\pm 2.07$\\
 2018-06-29 02:11 & 58298.09 & 13.79 & 596 & $0.61\pm 0.80$ & $0.12\pm 1.35$ & $0.04\pm 2.34$\\
 2018-06-29 21:18 & 58298.89 & 14.59 & 596 & $0.43\pm 0.48$ & $0.40\pm 1.27$ & $0.00\pm 1.90$\\
 2018-06-30 01:35 & 58299.07 & 14.77 & 598 & $0.90\pm 0.68$ & $0.00\pm 1.39$ & $0.74\pm 2.15$\\
 2018-07-02 21:18 & 58301.89 & 17.59 & 599 & $0.29\pm 0.56$ & $1.63\pm 1.41$ & $0.00\pm 2.28$\\
 2018-07-04 21:22 & 58303.89 & 19.59 & 597 & $1.60\pm 0.94$ & $2.73\pm 2.13$ & $0.43\pm 3.40$\\
 2018-07-06 21:25 & 58305.89 & 21.59 & 597 & $0.01\pm 0.71$ & $2.41\pm 1.52$ & $2.80\pm 2.52$\\
 2018-07-08 21:29 & 58307.90 & 23.60 & 597 & $0.00\pm 1.37$ & $1.00\pm 3.08$ & $0.00\pm 4.93$\\
 2018-07-10 21:17 & 58309.89 & 25.59 & 598 & $2.30\pm 1.84$ & $0.00\pm 3.33$ & $0.00\pm 6.11$\\
 2018-07-12 21:18 & 58311.89 & 27.59 & 597 & $0.98\pm 1.31$ & $2.42\pm 2.89$ & $0.00\pm 3.97$\\
 2018-07-14 21:15 & 58313.89 & 29.59 & 597 & $1.24\pm 1.32$ & $2.81\pm 2.62$ & $0.00\pm 4.70$\\
 2018-07-28 21:05 & 58320.88 & 36.58 & 595 & $0.00\pm 2.95$ & $0.00\pm 4.51$ & $0.00\pm 2.83$\\
\hline
\end{tabular}\\
$^{\dagger}$ Relative to the estimated explosion date of MJD58284.3 \citep{2018apj...865l...3p}.\\
$^{\ddagger}$ Measurements whose error is larger than the level of polarization (after correction for polarization bias) are considered to be zero.\\
\end{table*}

All {\it RINGO3} observations of AT2018cow were retrieved from the Liverpool Telescope Data Archive, having been reduced through the standard pipeline \citep{arnoldphd}.  The analysis of the data, for the derivation of the Stokes parameters, followed the procedures presented by  \citet{jermakphd}, \citet{2016mnras.458..759s}, \citet{2020apj...892...97j}, \citet{2021mnras.505.2662j} and \citet{2021mnras.503..312m}.  Aperture photometry of all sources in the field, and for the polarimetric calibration stars, was conducted using the Photutils package \citep{2020zndo...4049061B}.  A standard circular aperture, with radius of 5 pixels, was used throughout and the derived Stokes parameters were found to be generally insensitive to the size of the aperture with radius in the range $3 - 15$ pixels.  At later epochs ($>20\,\mathrm{days}$), however, the rapid decline in the brightness of AT~2018cow and contamination from the host galaxy created spurious (although not statistically significant, i.e. $<3\sigma$) polarization.

\begin{figure}
\includegraphics[width=8.5cm]{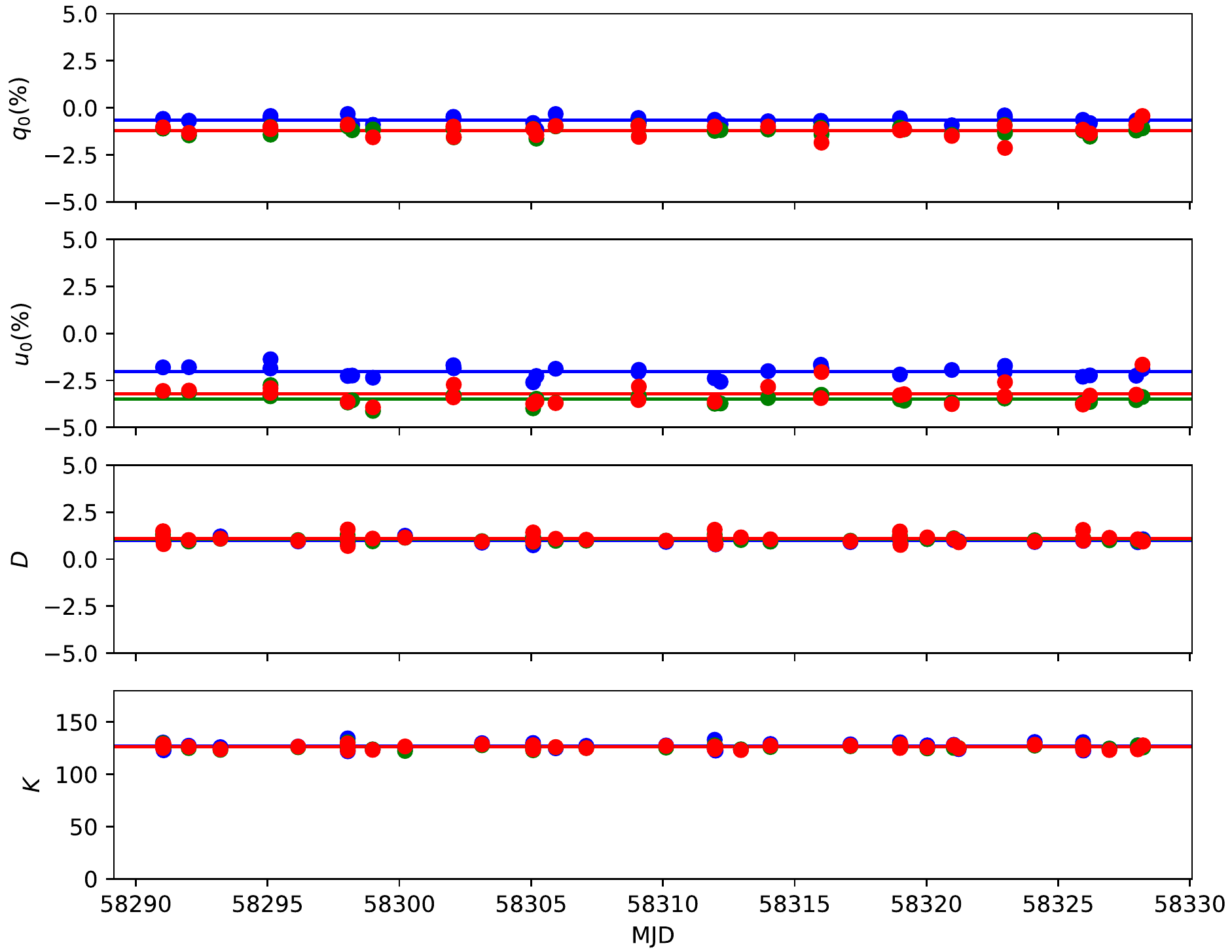}
\caption{The instrumental response of {\it RINGO3}, as measured using calibration observations of unpolarized and polarized standards over the period of the observations of AT~2018cow, in the {\it RINGO3} $b^{\ast}$ (blue circles), $g^{\ast}$ (green circles) and $r^{\ast}$ (red circles) bands (see Table \ref{tab:obs:calibration}).}
\label{fig:obs:calibration}
\end{figure}

Corrections for the baseline instrumental polarization ($q_{0}$ and $u_{0}$), the degree of instrumental depolarization ($D$) and the rotation of the instrument (with respect to the Celestial coordinate system; $K$) were calculated using observations of zero- and highly-polarized standard stars \citep{1992aj....104.1563s}.  The stability of the instrumental polarization properties is shown on Fig. \ref{fig:obs:calibration}.  Although there are variations in the response of {\it RINGO3}, from night to night, these are small (few tenths of $\%$) compared to the uncertainties of the polarization of AT~2018cow (see below) and the stars in the surrounding field (see Table \ref{tab:obs:calibration}).  The polarization $p$ was corrected for bias with the maximum probability estimator \citep{1985a&a...142..100s} using the expression provided by \citet{1997apj...476l..27w}.

A rudimentary photometric calibration of the Stokes $I$ fluxes derived from the {\it RINGO3} observations, to the approximately equivalent filters of the Sloan Digital Sky Survey (SDSS), was calculated using stars in the field around AT~2018cow that are also present in the SDSS photometric catalogue 12 \citep{2015apjs..219...12a}.  This calibration is relatively crude, since the colour correction term is highly uncertain for an object as blue as AT~2018cow given the significantly redder colours of the nearby stars; this is compounded by the non-standard transmission functions of the 3 {\it RINGO3} wavelength channels, defined by dichroic wavelength cutoffs, compared to the Sloan filter system.

\begin{table}
\caption{The average calibration properties of {\it RINGO3} (in parentheses are the standard deviation for each quantity) derived over the period of the observations of AT~2018cow. \label{tab:obs:calibration}}
\begin{tabular}{lcccc}
\hline
Channel & $q_{0} (\%)$ & $u_{0} (\%)$ & $D$ & $K(^{\circ})$ \\
\hline
$b^{\ast}$ & -0.67 (0.20) & -2.03 (0.29) & 0.99 (0.11) & 127.0 (2.9) \\
$g^{\ast}$ & -1.22 (0.20) & -3.48 (0.27) & 1.05 (0.11) & 126.2 (1.7) \\
$r^{\ast}$ & -1.21 (0.34) & -3.21 (0.33) & 1.09 (0.22) & 126.2 (1.9) \\
\hline
\end{tabular}
\end{table}

Transients can exhibit significant changes in brightness over the course of their evolution.  The interpretation of polarimetry, being dependent on calculating small differences between measured intensities, is particularly sensitive to changes in the levels of signal-to-noise associated with the change in brightness of the target transient. In the case of AT~2018cow, its significant drop in brightness between the epoch of the first and last {\it RINGO3} observation \citep[corresponding to $\Delta m_{V} \approx 4.5 \,\mathrm{mags}$; see ][]{2019mnras.484.1031p}, makes comparing early and late-time polarization behaviour difficult.  We have also rebinned the observed data in time, following the prescription of \citet{2013natur.504..119m},  to achieve constant levels of signal-to-noise (corresponding to constant uncertainties of $0.5$, $1$ and $1.75\%$, in the $b^{\ast}$, $g^{\ast}$ and $r^{\ast}$ bands, respectively) to match the signal-to-noise of the first two nights of observations at the expense of time resolution at the later epochs.

%%%%%%%%%%%%%%%%%%%%%%%%%%%%%%%%%%%%%%%%%%%%%%%%%%%%%%%%%%%%%%%%%%%%%%
% RESULTS & ANALYSIS - RESULTS & ANALYSIS - RESULTS & ANALYSIS
% RESULTS & ANALYSIS - RESULTS & ANALYSIS - RESULTS & ANALYSIS
% RESULTS & ANALYSIS - RESULTS & ANALYSIS - RESULTS & ANALYSIS
% RESULTS & ANALYSIS - RESULTS & ANALYSIS - RESULTS & ANALYSIS
%%%%%%%%%%%%%%%%%%%%%%%%%%%%%%%%%%%%%%%%%%%%%%%%%%%%%%%%%%%%%%%%%%%%%%
\section{Results \& Analysis}
\label{sec:res}

\begin{figure*}
\includegraphics[width=17cm]{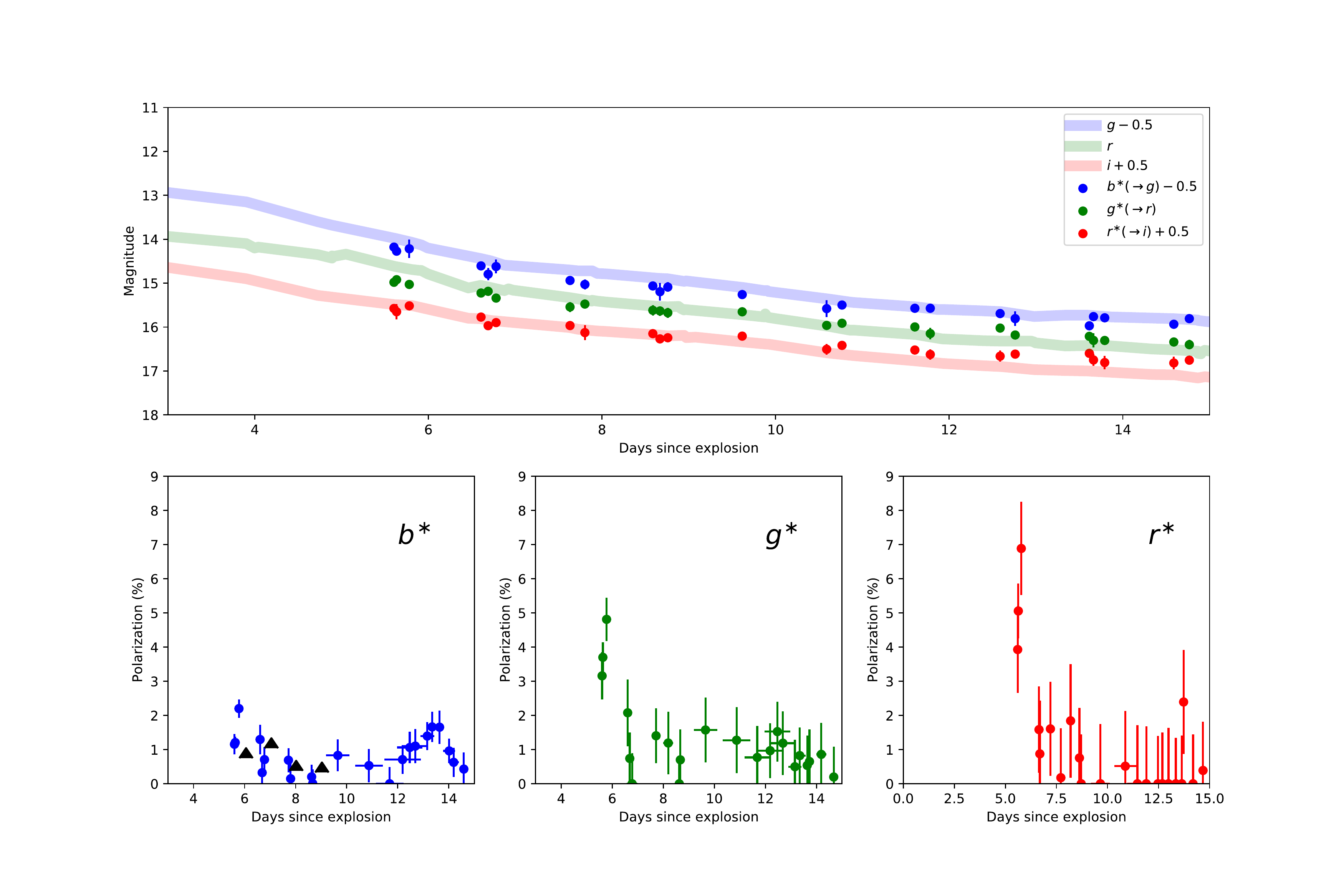}
\caption{The observed evolution of the polarization of AT2018cow measured in the {\it RINGO3} $b^{\ast}$ (blue), $g^{\ast}$ (green) and $r^{\ast}$ (red) bands.  {\it Top panel)} The lightcurve of AT2018cow \citep{2019mnras.484.1031p} (solid lines) and overlaid are photometry derived from the {\it RINGO3} observations transformed to the approximately equivalent filters of the Sloan Digital Sky Survey.  {\it Bottom panels)} The polarization of AT~2018cow measured by {\it RINGO3} in the $b^{\ast}$, $g^{\ast}$ and $r^{\ast}$ bands, for which the temporal binning of the data has been chosen to fix the precision of each binned data point (see text).  The black triangles are observations of AT2018cow acquired with the 2.3m Bok telescope \citep{2018atel11789....1s}, approximately consistent with the {\it RINGO3} $b^{\ast}$ band (the reported uncertainties are smaller than the size of the points)}
\label{fig:res:lcpol}
\end{figure*}

Due to the large, multi-wavelength dataset of polarimetry of AT~2018cow, unlike previous studies it is more useful to explore the evolution of the polarization with time rather than Stokes $q$ and $u$.  The polarization measured for AT~2018cow is shown in Fig. \ref{fig:res:lcpol} (rebinned to constant signal-to-noise) and the raw polarization measurements are presented in Table \ref{tab:obs:pol}.  Significant levels ($>3\sigma$) of polarization are observed in all 3 {\it RINGO3} channels in 3 independent observations on the first night, corresponding to $\sim 5.7\,\mathrm{days}$ post-explosion or 2.8 days after maximum-light \citep{2018apj...865l...3p, 2019mnras.484.1031p}. The rate of increase in the observed polarization on the first night corresponds to $\mathrm{d}p/\mathrm{d}t = 0.26\pm0.03$, $0.34\pm0.07$ and $0.60\pm0.19\,\%\,\mathrm{per\,hour}$, in the $b^{\ast}$, $g^{\ast}$ and $r^{\ast}$ bands, respectively. Also shown in Figure \ref{fig:res:lcpol} are polarization measurements of AT2018cow acquired with the 2.3m Bok telescope, for which median values over the wavelength range $\mathrm{4300-7400\mathring{A}}$ (which is approximately consistent with the {\it RINGO3} $b^{\ast}$ band) were reported by \citet{2018atel11789....1s}.

\subsection{Significance of the detection of early polarization}

As {\it RINGO3} is a single channel polarimeter, it suffers from a high systematic floor which can hinder attempts to measure low levels of polarization ($<2\%$), but make it excellent for monitoring the rapid evolution of objects with high polarization \citep{2013natur.504..119m}.  We consider three separate tests to establish the significance of the detections of polarization of AT~2018cow at early-times \citep{2009natur.462..767s}: a) the noise model of each observation, characterised by the polarization uncertainty as a function of the brightness of AT~2018cow and other stellar sources in the field; b) the correlation of the polarization measurements with the moon illumination and distance of the moon from the position of AT~2018cow; and c) similarities of the data to other reported polarimetric observations.

\subsubsection{The polarization of AT~2018cow with respect to other stars in the {\it RINGO3 field}}
\label{sec:analysis:significance:surround}
In considering the significance of the polarization of AT~2018cow, we can establish whether it is anomalous or typical of the {\it RINGO3} observations, given the observing conditions, through comparison with other stars in the field.  For low-polarization or unpolarized sources, larger degrees of polarization (with commensurate large uncertainties) will be inferred for fainter targets.  A trend is expected, therefore, that the apparent degree of polarization will increase (along with the uncertainties) for fainter low polarization or unpolarized sources (even after correction for the polarization bias using the maximum probability estimator).  We use our aperture photometry of all sources in the field around and including AT~2018cow to derive an instrumental magnitude $m_{inst} = -2.5 \log_{10} f_{TOT}$, where $f_{TOT}$ is the sum of all photometry, acquired at each {\it RINGO3} rotor position, for each source. We conducted this analysis for all three {\it RINGO3} channels.

\begin{figure*}
  \includegraphics[width=14cm]{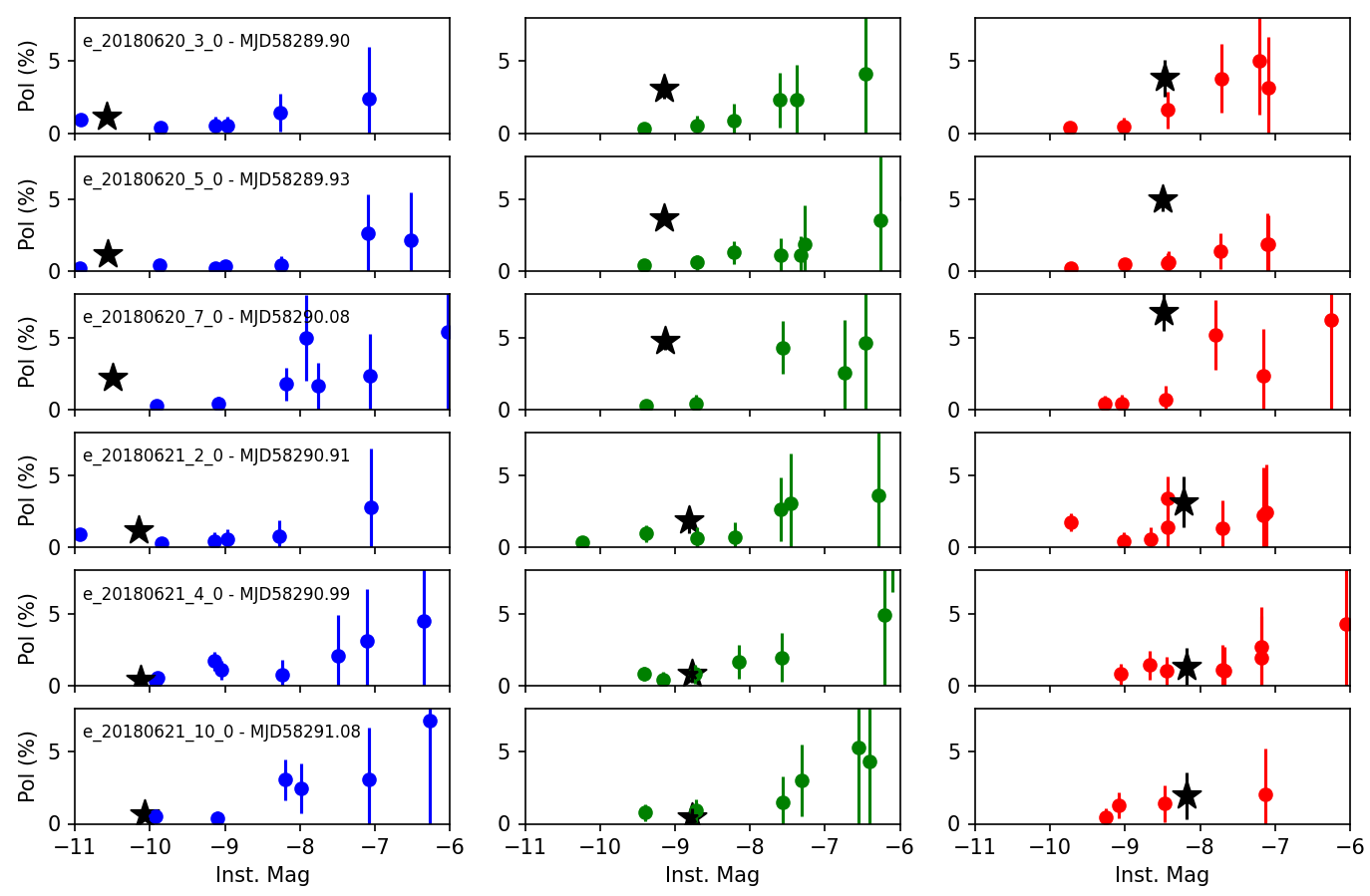}
\caption{The polarization of sources in the field surrounding AT~2018cow in the $b^{\ast}$ (left), $g^{\ast}$ (middle) and $r^{\ast}$ (right) bands for the first six epochs (corresponding to the first two nights of observations).  The polarization values have been corrected for bias, using the expression of \citet{1997apj...476l..27w}.  In each panel AT2018cow is indicated by the $\star$ symbol.  For the first 3 epochs, the polarization of AT~2018cow exhibits an excess (particularly in the $g^{\ast}$ and $r^{\ast}$ bands) compared to other sources in the field of similar brightness.}
\label{fig:analysis:significance:surrounding:surrounding}
\end{figure*}

As we only require a relative photometric calibration, it is not necessary to derive zeropoints to place the photometry on a proper photometric scale.  Although we have computed an approximate photometric calibration (see Fig. \ref{fig:res:lcpol} and Section \ref{sec:obs}), the instrumental magnitudes provide a better measure of the relative brightness of sources in the field of view in the specific wavelength bands of {\it RINGO3}.  This does mean, however, that the photometric magnitude scale may differ between the wavelength channels and between epochs.  In Fig. \ref{fig:analysis:significance:surrounding:surrounding}, we show the resulting measured brightness and polarization for all stellar sources, including AT~2018cow, at the first six epochs (covering the first and second nights of observations).  The plots show clearly the higher and increasing degree of polarization of AT~2018cow, compared to other sources in the field of view with similar brightness in the first three observations.  In the following night, the polarization properties of AT~2018cow, when we no longer see significant polarization, are consistent with the surrounding stars; such that the early polarization is clearly anomalous and intrinsic to AT~2018cow, rather than a property of the observations.

\subsubsection{Correlation of the polarization of AT~2018cow with lunar illumination and distance}
\label{sec:analysis:significance:moon}
\begin{figure*}
    \centering
    \includegraphics[width=15cm]{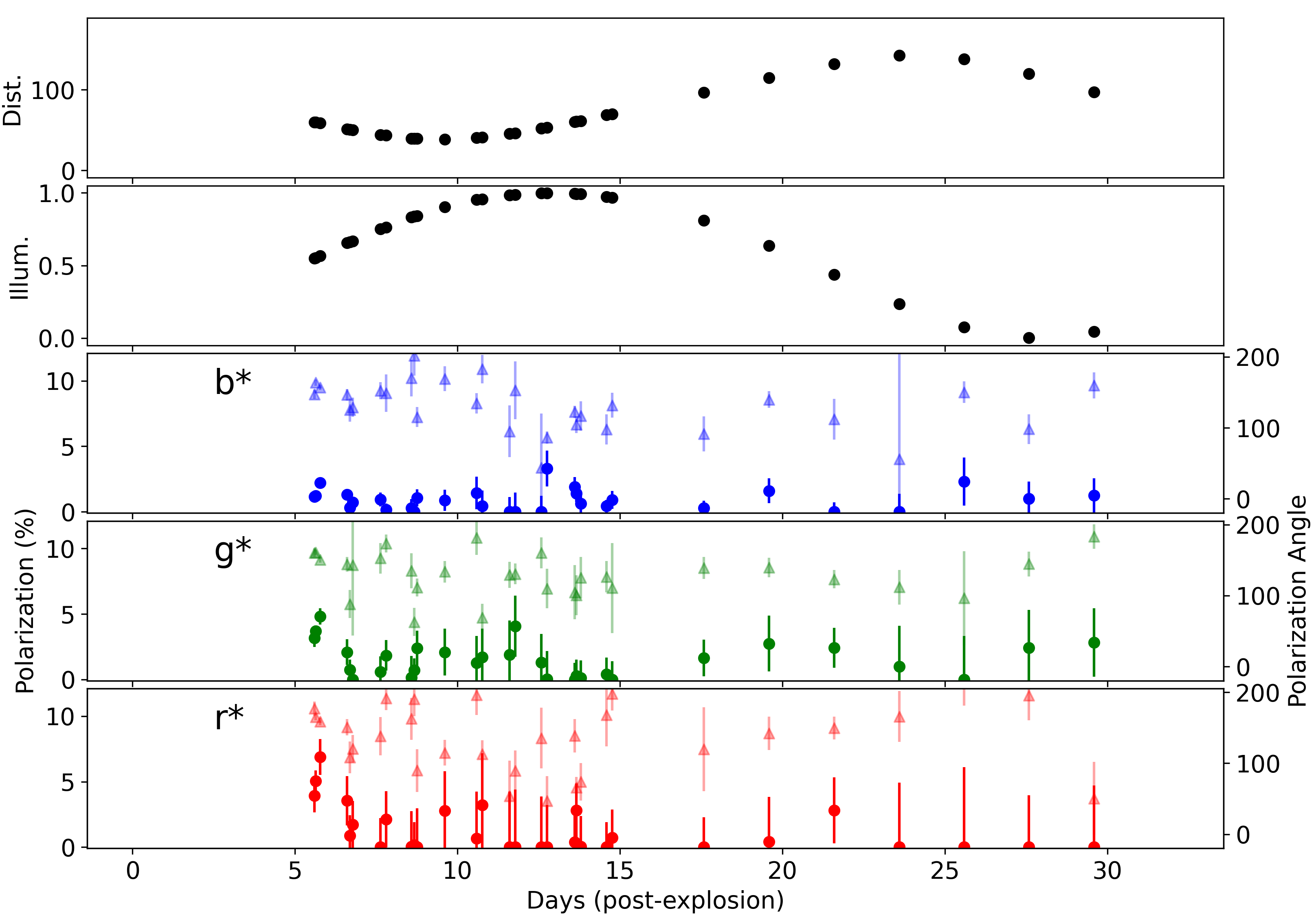}
    \caption{The evolution of the degree of polarization ($\bullet$) and polarization angle ($\blacktriangle$) of AT2018cow, in the three RINGO3 bands $b^{\ast}/g^{\ast}/r^{\ast}$ (lower panels) with respect to changing moon distance (top panel) and illumination (second panel) over the period of observations.  }
    \label{fig:analysis:significance:moon:moon}
\end{figure*}

Scattered moonlight not only increases the brightness of the sky background in astronomical observations (increasing the level of noise associated with photometric measurements) but may also be polarized itself, contributing an additional signal in the polarization analysis.  We compared our observed levels of polarization with lunar illumination (full moon corresponding to an illumination of 1 and a new moon corresponding to dark conditions and an illumination of 0) and the distance of the moon (in degrees) from the position of AT2018cow.  As can be seen from Figure \ref{fig:analysis:significance:moon:moon}, the first epoch is taken under average conditions.  Later, at $\sim 9\,\mathrm{days}$ (or MJD58293.92) when the lunar distance is at a minimum and the illumination is largest,  we do not see any corresponding increase in the degree of polarization (when the effect of the spurious polarization arising from the moon is expected to be at its worst).  This reflects previous reports that the lunar illumination and distance do not significantly affect the performance of the {\it RINGO3} polarimeter \citep{2016mnras.458..759s} and spurious polarization due to contamination by moonlight is not the cause for the early observed polarization of AT~2018cow.

\subsubsection{Comparison of {\it RINGO3} measurements with other reported polarization measurements}
\label{sec:analysis:significance:opol}
\begin{figure}
    \centering
    \includegraphics[width=8.5cm]{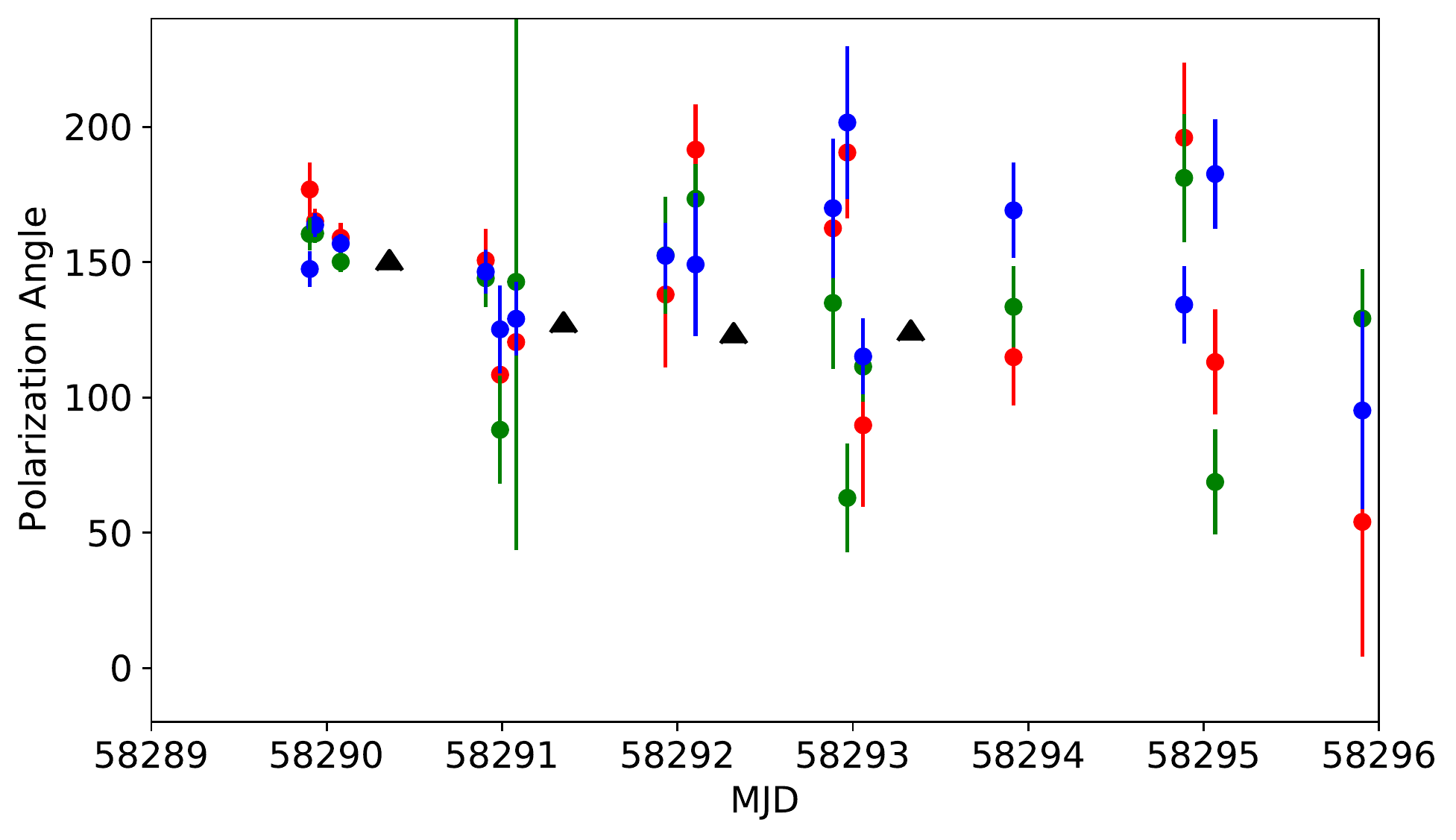}
    \caption{Evolution of the polarization angle of AT2018cow.
    The black triangles are measurements made with the 2.3m Bok Telescope and are most directly comparable with the measurements made in the {\it RINGO3} $b^{\ast}$ channel.  It is important to note that the increased scatter in the polarization angle, measured by {\it RINGO3} at later times, is correlated with the decrease in both the brightness and polarization of AT~2018cow.}
    \label{fig:analysis:significance:angle:angle}
\end{figure}

In addition to our {\it RINGO3} polarimetric observations of AT~2018cow, later polarimetric observations, conducted by the 2.3m Bok Telescope, have also been reported \citep{2018atel11789....1s}.  The observed evolution of the degree of polarization and polarization angles matches our observations of the decreasing level of polarization observed after 6 days (see Figs. \ref{fig:res:lcpol} and \ref{fig:analysis:significance:angle:angle}).  The Bok Telescope observations concur with our derived polarization degree and polarization angle and there is also a report of a similar wavelength dependence in the degree of polarization, reaching $\sim 2\%$ at $7550\mathrm{\mathring{A}}$ (at $\sim 6.7\,\mathrm{days}$) in agreement with our observations, on the same night, in the $g^{\ast}$ and $r^{\ast}$ bands.
At later epochs, as AT~2018cow fades and the polarization decreases, the scatter in the polarization angle derived from the {\it RINGO3} observations increases as the uncertainty on the polarization angle $\Delta \chi \sim \Delta p / p$.  We note that between our first and last {\it RINGO3} observations AT~2018cow had dropped in brightness by a factor of $\sim 60$ in the $V$-band \citep{2019mnras.484.1031p}, which severely limits constraints on its late-time polarimetric behaviour for low levels of polarization given the relatively high systematic floor of the {\it RINGO3} polarimeter and increasing contamination by the host galaxy.

\subsection{Interstellar Polarization}
\label{sec:analysis:isp}

The interstellar polarization (ISP) is expected to be a constant additional source of polarization, arising due to dichroic absorption by aligned dust grains in both the Milky Way and the host galaxy.  The reddening towards AT2018cow has been previously assumed to be low, and dominated by the Milky Way \citep{2018apj...865l...3p,2019mnras.484.1031p} with a value of $E(B-V) = 0.07 \-- 0.08\,\mathrm{mags}$ \citep{2011apj...737..103s}.  This corresponds to an upper limit on the degree of polarization $p_{ISP} < 9 \times E(B-V) = 0.7\%$ \citep{1975apj...196..261s}.

From the Heiles catalogue \citep{2000aj....119..923h}, we find there is one Galactic star (HD147266) within 2 degrees of the line of sight towards AT~2018cow, located at only a distance of $\sim 100\,\mathrm{pc}$  \citep{2021aj....161..147b}.  As such, this star does not sample the full column of Galactic dust towards AT~2018cow; however, its low polarization of $0.16\%$ with a polarization angle of $\mathrm{134^{\circ}}$ may be indicative of the scales of polarization we might expect to be associated with any ISP component arising in the Galaxy.

\begin{figure}
    \centering
    \includegraphics[width=8.5cm]{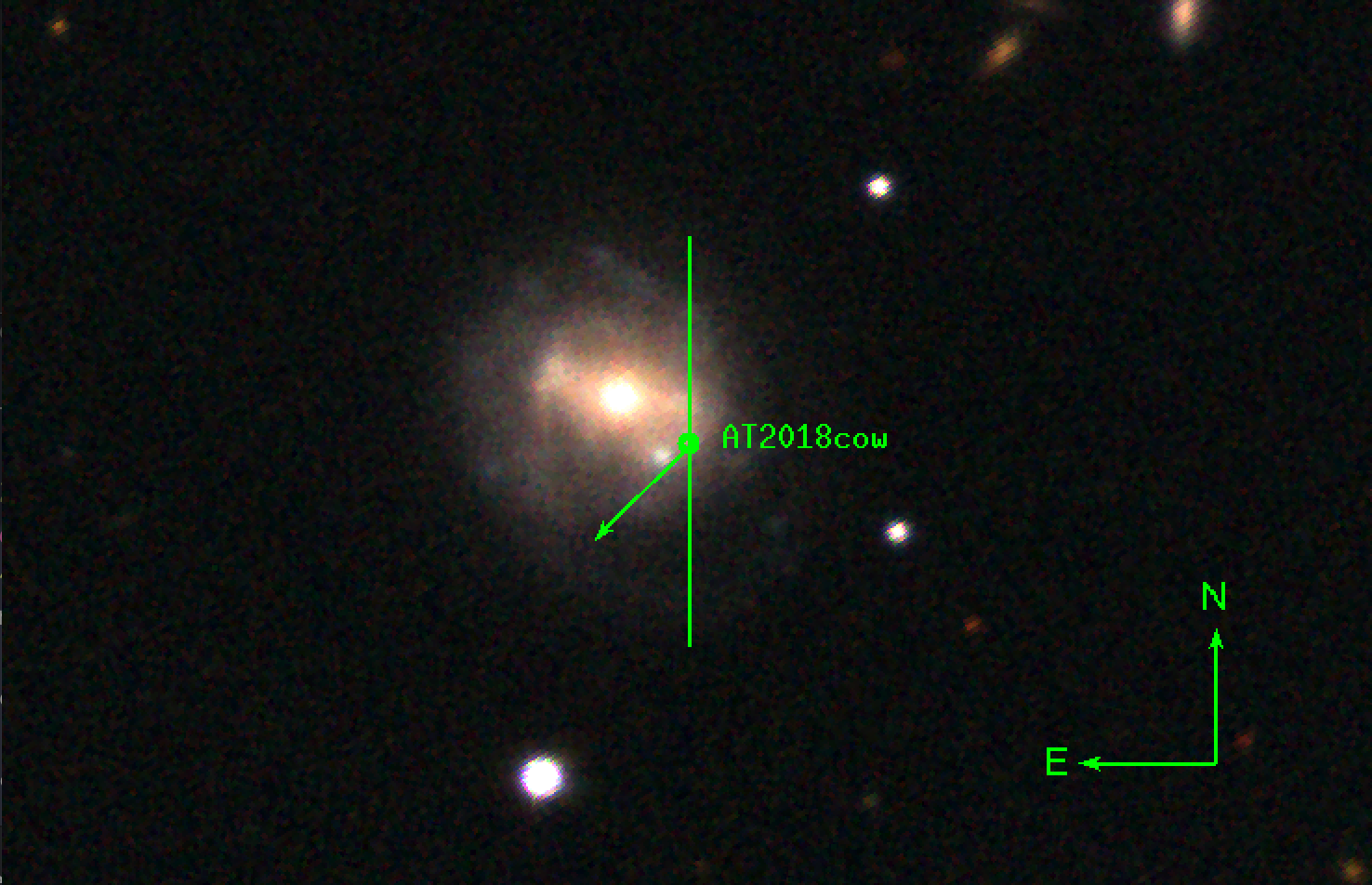}
    \caption{Sloan Digital Sky Survey $g^{\prime}r^{\prime}i^{\prime}$ image of the location of AT~2018cow in the galaxy CGCG137-068.  The position of the transient is indicated by the green circle.  The approximate alignment of the spiral arm containing the location of the transient, with a position angle of $\sim 135^{\circ}$ measured East of North, is shown relative to the vertical.}
    \label{fig:analysis:isp:align}
\end{figure}

We see from our early time observations (days $5 \-- 7\,\mathrm{days}$) that the polarization angle evolves at the same time as the degree of polarization (see Figs. \ref{fig:res:lcpol} and \ref{fig:analysis:significance:angle:angle}), and a similar evolution is observed for the reported observations from the Bok telescope.  Although we do not derive the ISP directly from the {\it RINGO3} observations, a significant change in the polarization on such short time-scales must be intrinsic to the transient.  The lower levels of polarization reported for the observations with the Bok telescope \citep{2018atel11789....1s} are close to the systematic floor of our {\it RINGO3} observations, suggesting any ISP present is small ($<1\%$) and significantly smaller than the significant polarization we have observed at $\sim 5.7\,\mathrm{days}$. At the same time, we observe a rotation of the polarization angle of $-30^{\circ}$ between the first night and subsequent observations (also seen in the Bok observations - see Fig. \ref{fig:analysis:significance:angle:angle}). If the latter data are consistent with null intrinsic polarization, then the later polarization angles observed both by {\it RINGO3} and the Bok telescope are consistent with a plausible Galactic ISP component. The polarization angle may also be consistent with the expected alignment of dust grains in the host galaxy (CGCG137-068) being parallel to the spiral arms at the location of the transient \citep{1987mnras.224..299s} at $\sim 135^{\circ}$ (see Fig. \ref{fig:analysis:isp:align}).

%%%%%%%%%%%%%%%%%%%%%%%%%%%%%%%%%%%%%%%%%%%%%%%%%%%%%%%%%%
% DISCUSSION & CONCLUSIONS - DISCUSSION & CONCLUSIONS
% DISCUSSION & CONCLUSIONS - DISCUSSION & CONCLUSIONS
% DISCUSSION & CONCLUSIONS - DISCUSSION & CONCLUSIONS
% DISCUSSION & CONCLUSIONS - DISCUSSION & CONCLUSIONS
%%%%%%%%%%%%%%%%%%%%%%%%%%%%%%%%%%%%%%%%%%%%%%%%%%%%%%%%%%

\section{Discussion \& Conclusions}
\label{sec:discussion}

%\subsection{The significance of the observed high degree of polarization}
In the absence of a relativistic or highly magnetized flow \citep{2020mnras.491.4735b,2019apj...871...73h} such as for Gamma Ray Bursts \citep{2007sci...315.1822m}, polarization that is intrinsic to the transient requires a geometric configuration for the thermal electron scattering atmosphere that is capable of producing a high degree of polarization.  The polarization measured on the first night is greatly in excess of the theoretical maximum polarization limit for an oblate spheroidal, asymmetric explosion ($\sim 4\%$) for electron scattering dominated atmospheres \citep{1991a&a...246..481h} and in excess of the levels of intrinsic polarization previously observed for non-relativistic SN explosions \citep{2008ara&a..46..433w}, including the previous record-holder: the Type IIn SN~2017hcc \citep[with continuum polarization $p_{V} = 4.84\pm0.02\%$, with little expected ISP contribution; ][]{2017atel10911....1m}.  Such high levels of polarization are also inconsistent with a prolate spheroidal structure \citep{1980a&a....86..198d}, excluding the origin of the polarization in an extended jet-like structure.

\begin{figure}
\includegraphics[width=8.5cm]{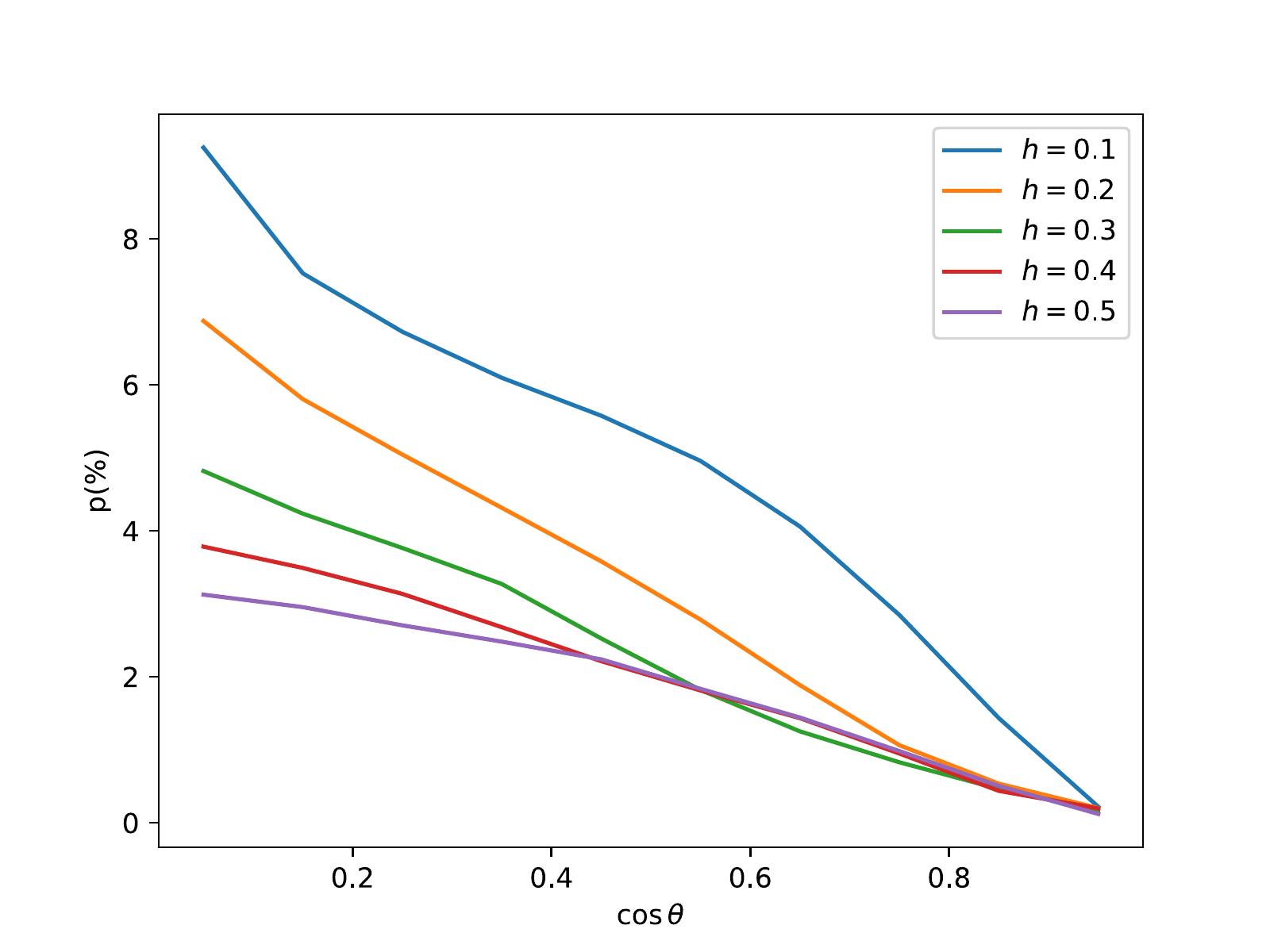}
\caption{The predicted polarization as a function of the inclination angle $\theta$ for models of flat disks with optical depth of $\tau = 5.0$ and thickness $h$ (relative to the projected maximum extent of the disk on the sky).}
\label{fig:diskinclination}
\end{figure}

Higher levels of polarization can be produced by other geometric configurations, beyond those that are part of the family of spheroidal models.  A disk-like configuration in which geometry is governed by the thickness of the disk can yield higher polarizations.  We utilised a Monte Carlo Radiative Transfer simulation to consider electron scattering in a simple disk, in which the density of electrons was allowed to scale with the radius according to $r^{-2}$ and the disk having constant thickness $h$ relative to the maximum projected extent of the disk on the sky.  The total optical depth to electron scattering, in the radial direction from the inner boundary of the simulation volume (corresponding to inner edge of the disk) to the outer edge of the disk, was kept fixed with $\tau = 5.0$.  Photon packets were inserted at the inner boundary of the simulations, with all photons undergoing an initial non-polarizing isotropic scattering event before their trajectories were followed through the disk.  Photon packets that, once emitted, crossed the inner edge of disk were assumed to be reabsorbed and a new photon packet was re-emitted.  The resulting polarization, as a function of the cosine of the inclination angle at which the disk might be observed, for different values of the disk thickness $h$, is shown in Fig. \ref{fig:diskinclination}.  It can be seen that for very flat disks, with $h \lesssim 10\%$, values of the polarization are produced similar to those observed for AT~2018cow on the first night of {\it RINGO3} observations.  A disk-like configuration will yield maximum polarization when observed edge-on, however high-levels of polarization may still be perceived when viewing the disk within $\pm 30^{\circ}$ of edge-on; which means that such a configuration does not require a particularly special orientation for high polarization to be apparent.

%\subsection{Wavelength dependence of the polarization}
Thomson scattering is a wavelength independent process, and so the observed wavelength dependence of the polarization would require an additional process to effectively depolarize or repolarize the light (if it does not arise for a non-thermal process). Continuum polarization, produced by Thomson scattering, may acquire a wavelength dependence through three processes: dilution, depolarization or dust.  As discussed in Section \ref{sec:analysis:isp}, the rapid time variability of the polarization observed in the first nights of the {\it RINGO3} observations implies that the origin of the polarization is not due to dichroic absorption of aligned dust grains.  Depolarization would require subsequent reprocessing of previously polarized photons, effectively erasing their polarization \citep[e.g. as is seen in the emission line components of P Cygni profiles; ][]{1984mnras.210..829m}, while dilution would require the presence of an additional, unpolarized flux component to reduce the overall proportion of polarized flux.

Line blanketing over particular wavelength ranges, associated in particular with Fe at $\mathrm{\lesssim\,5000\mathring{A}}$, may result in depolarization due to multiple line interactions; and this has been observed in Type Ia SNe \citep{2001apj...556..302h}.  In the case of AT~2018cow, a broad feature was seen to emerge at $\sim 4$ days at $\sim 4600\,\mathring{A}$, with a Full Width at Half-Maximum of $\mathrm{\sim 1500\,\mathring{A}}$, which disappeared by day 8 \citep{2019mnras.484.1031p}.  Such a transient feature has been observed in other 2018cow-like FBOTS, such as AT~2020mrf where it was similarly observed to appear at 4.8 days \citep{2022apj...934..104y}.  \citet{2019mnras.484.1031p} notes that there was initially some resemblance to Fe {\sc ii} as seen in broad-line Type Ic SNe, that might be associated with Gamma Ray Bursts; however, the disappearance of this feature and the lack of other emerging line features at early times casts significant doubt on this identification.  If this transient feature is depolarizing, then this could explain the modest polarization observed on the first night of {\it RINGO3} observations seen in the $b^{\ast}$-band compared to the redder wavelength ranges.  In this case, the high polarization observed in the $r^{\ast}$-band would reflect the true continuum polarization.

\begin{figure}
    \centering
    \includegraphics[width=8.5cm]{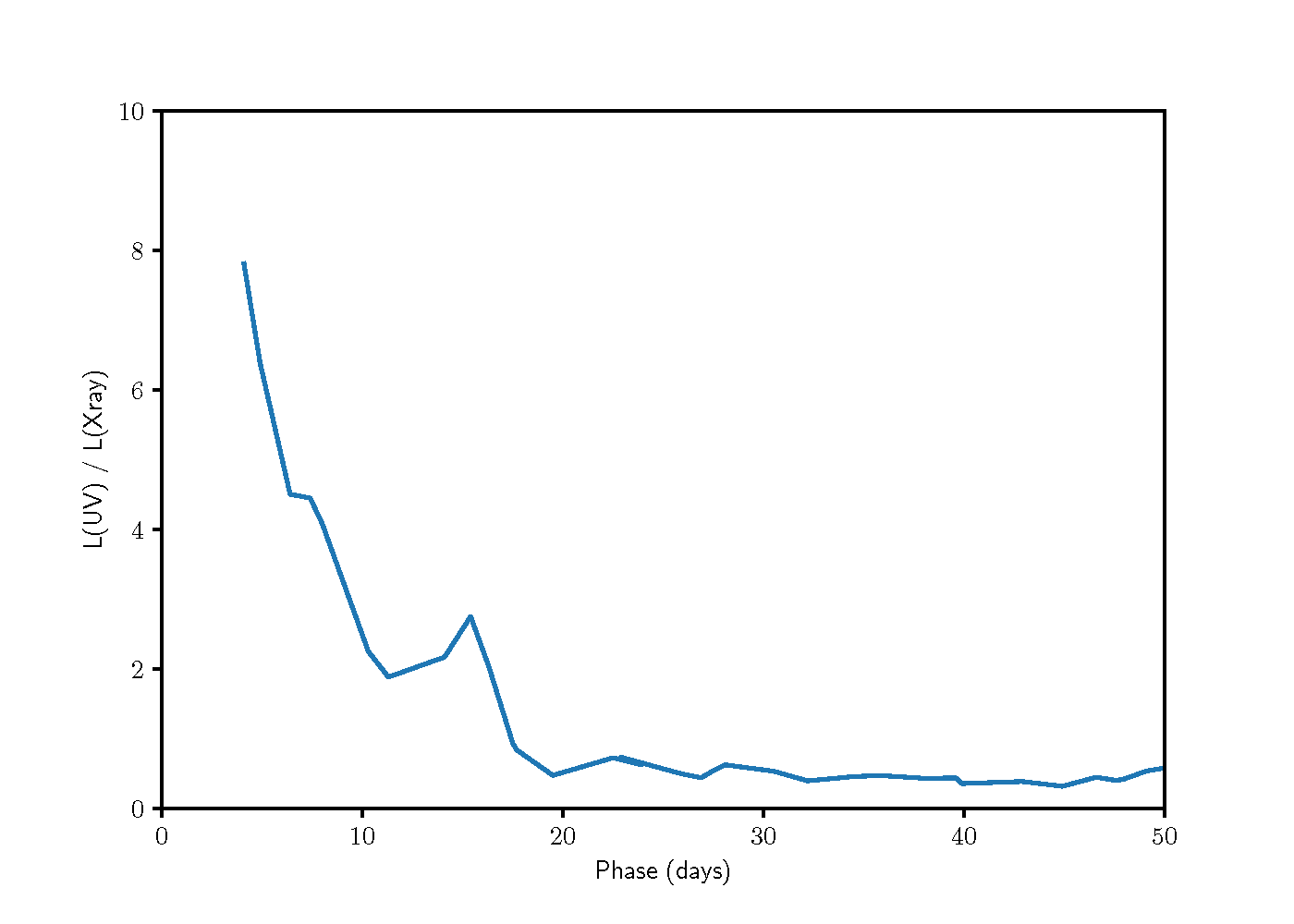}
    \caption{The ratio of UV to X-ray luminosity of AT~2018cow as a function of time, as measured by the Swift UVOT and XRT \citep{2019mnras.484.1031p, 2019apj...872...18m}.}
    \label{fig:discussion:uvxray}
\end{figure}

\citet{2019apj...872...18m} propose that downscattering of the significant X-ray flux (which itself is in excess of that expected given the radio synchrotron emission) may be a significant contributor to UV-optical spectrum. The ratio of the UV to X-ray luminosity indicates the degree to which X-rays may be being down-scattered through Compton scattering to UV energies.  The evolution of the ratio of UV to X-ray luminosity was calculated from observations of AT~2018cow conducted by the Neil Gehrels Swift Observatory, with the Ultraviolet/Optical Telescope (UVOT) \citep{2019apj...871...73h} and X-ray Telescope (XRT) \citep{2019apj...872...18m}. The flux densities measured from Swift UVOT observations in the $UVW2$, $UVM2$, $UVW1$ and $U$ bands were used to characterise a ``total" UV flux, by numerically integrating across the UV spectral energy distribution covering the wavelength range $\mathrm{2140 - 3493\,\mathring{A}}$.  The data were corrected for Galactic extinction \citep{1999pasp..111...63f} assuming a value for the foreground reddening of $E(B-V) = 0.07\,\mathrm{mags}$.  The unabsorbed X-ray flux, in the energy range $0.3 - 10\,\mathrm{keV}$, was used to characterise the X-ray luminosity. Due to the sparse nature of the X-ray observations, relative to the UVOT observations, Gaussian processes \citep{2015itpam..38..252a} were used to estimate the X-ray flux at the epochs corresponding to the UV observations.  The calculated ratio of the UV to X-ray luminosities is shown on Fig. \ref{fig:discussion:uvxray} and clearly shows the UV flux dominated at epochs $<20\,\mathrm{days}$ post-explosion.  Depending on the wavelength dependence of the contribution of downscattered X-rays to the optical spectrum, which may be assumed to be stronger at bluer wavelengths, the optical spectrum and wavelength range covered by the {\it RINGO3} may be a composite, for which we hypothesise the redder wavelengths may be truly reflective of the polarization produced by an asymmetric electron scattering atmosphere, while the polarization observed at bluer wavelengths could have been diluted by a contribution arising from downscattered X-rays.

It is difficult to ascertain the correlation of the early polarization spike with other behaviour observed for AT~2018cow because: 1) the first multicolour optical photometry was only acquired $<3$ days before the first polarimetric observation was conducted; and 2) there are multiple timescales over which the behaviour has changed which, due to the rapid evolution of AT~2018cow, occured over a short period of time.  The UV luminosity dominates over the X-ray luminosity until $\sim 20$ days (as shown Fig. \ref{fig:discussion:uvxray}), which coincides with the transition of the X-ray luminosity from a plateau to a decline with increased variability \citep{2019apj...871...73h, 2019apj...872...18m} and the emergence of H and He emission lines in the optical spectrum \citep{2018apj...865l...3p}.  The rise time for AT~2018cow was only $\sim 3\,\mathrm{days}$, the X-ray lightcurve exhibits an initial decline before settling on the plateau at $\sim 4\,\mathrm{days}$ \citep{2019apj...872...18m} and the UV lightcurve exhibited a change in slope at $\sim 9\,\mathrm{days}$ \citep{2018apj...865l...3p}.  We note that, based on the photometry tabulated by \citet{2019mnras.484.1031p}, there is no significant change in optical colour at the time the polarization spike was observed.  It is not clear, therefore, that there is any other particular change in the evolution of AT~2018cow that is necessarily and specifically correlated with the behaviour observed for the polarization at optical wavelengths.

At $\approx 13\,\mathrm{days}$, another increase in the polarization is apparent in the $b^{\ast}$-band polarization lightcurve at a level of $\sim 2\%$.   Unlike the initial polarization spike, this feature is not perceptible in the observations at other wavelengths (although the decrease in signal-to-noise makes it difficult to detect significant polarization at these wavelengths to a similar precision as the $b^{\ast}$-band). Similar conditions, as described above, for inducing the possible wavelength dependence of the polarization at $\sim 6\,\mathrm{days}$ could also be in action at this time, although distinctly different to enhance the polarization at blue wavelengths.  An alternative explanation, given the presence of significant UV flux at this time, is that this polarization may constitute a reflection from the disk, that is distinct from unpolarized thermal emission that dominates at redder wavelengths.

%\subsection{Scenario Constraints}
The rapid rise and decline in the lightcurve of AT~2018cow presents a conundrum if this event is to be considered in the paradigm of core-collapse supernovae.  The rise time suggests progenitor dimensions similar to that of a red supergiant (RSG; $R \approx 10^{14}\,\mathrm{cm}$ or $10\,\mathrm{au}$), while the rapid decline is incompatible with a massive RSG envelope \citep{2019mnras.484.1031p,2019apj...872...18m,2019apj...871...73h}.  The early behaviour of the lightcurve favours the presence of a circumstellar medium that, unlike a massive envelope, is spatially confined with a limited radial extent \citep{2019apj...871...73h} or is not spherically symmetric.  The optical polarization presented here suggests that both factors may be at work, with the degree of optical polarization requiring a geometry akin to a disk with finite extent (we note that our simulation, presented above, only requires the disk to be sufficiently optically thick to light travelling through the disk, but does not have a requirement that it possess any specific spatial scale).  The limited spatial extent for such a disk-like configuration may also explain the brevity of the polarization spike observed at $5.7$ days.

\citet{2019mnras.484.1031p} draw comparisons with the evidence for material from recent mass loss, using ``flash spectroscopy", in very young supernovae \citep{2014natur.509..471g}.  Given the time at which the polarization pulse is observed, assuming homologous expansion, a spatial scale of $\sim 10^{14}\,\mathrm{cm}$ requires an ejecta velocity $\sim 6000\,\mathrm{km\,s^{-1}}$, which is similar to velocity of the slow equatorial ejecta proposed by \citet{2019apj...872...18m} and the velocity widths of the late-time optical emission lines.  \citet{2019apj...872...18m} suggest that shock breakout could be responsible for the early behavior of AT~2018cow, while a central engine is required at later epochs, and so the polarization spike could correspond to shock breakout through a disk that may be related to pre-explosion mass loss.  The lack of observed emission lines, as seen for SN~2013cu \citep{2014natur.509..471g}, may suggest that the disk is still optically thick to electron scattering ($\tau_{e} >> 1$) which may suppress line formation, until the disk is consumed by the ejecta.  The presence of such pre-existing disk-like material is also a feature for a number of models for the origin of AT~2018cow and luminous FBOTs, such as: the Common Envelope Jets Supernova \citep{2022raa....22e5010s}, a merger between a Wolf-Rayet star and a black hole \citep{2022apj...932...84m} and the merger of two white dwarfs \citep{2019mnras.487.5618l}.

%\subsection{Future}
The precision of the polarization measurements of AT~2018cow was limited by the sensitivity of {\it RINGO3} at low polarization levels, and that the brief period of high polarization was unexpected and the observations were not designed to specifically capture such an event.  Future observations of luminous FBOTs, at sufficiently early times, could be used to determine if spikes in polarization are a common property of these events, as other features (such as the light curve shape and the observation of a broad absorption/depression at blue wavelength) have been found to be.  The later {\it RINGO3} observations were hampered by the limited instrumental sensitivity, the rapid decline in the brightness of AT~2018cow and the presence of strong background contamination from the host galaxy.  In late 2020, {\it RINGO3} was replaced by the Multicolour OPTimised Optical Polarimeter \citep[MOPTOP; ][]{2020mnras.494.4676s}, a dual-beam polarimeter with a lower systematic floor, capable of achieving a polarization accuracy of $0.6\%$ for $m_{R} = 17\,\mathrm{mag}$ compared to $2.6\%$ for {\it RINGO3} (as evidenced in Fig. \ref{fig:res:lcpol} for the $r^{\ast}$-band).  The rapid response of the Liverpool Telscope to fast evolving transients, such as FBOTs, will be critical to capturing the apparently complex evolution of these objects at early times and, with MOPTOP, may provide direct evidence to the role of asymmetry in producing their peculiar behaviour \citep{2021mnras.503..312m}.

\section*{Acknowledgements}
JRM acknowledges support from the Science and Technologies Facilities Council (STFC) grant ST/V000853/1.  PAH  acknowledges support by the National Science Foundation (NSF) grant AST-1715133.   The research of Y.Y. is supported through a Bengier-Winslow-Robertson Fellowship.  KW acknowledges funding from the European Research Council (ERC) under the European Union Horizon 2020 research and innovation programme awarded to Prof A.~Levan (grant agreement no 725246) and from a UK Research and Innovation Future Leaders Fellowship awarded to Dr.~B. Simmons (MR/T044136/1). A. G. acknowledges the financial support from the Slovenian Research Agency (grants P1-0031, I0-0033, J1-8136, J1-2460). The Liverpool Telescope is operated on the island of La Palma by Liverpool John Moores University in the Spanish Observatorio del Roque de los Muchachos of the Instituto de Astrofisica de Canarias with financial support from the UK Science and Technology Facilities Council (STFC).  This research made use of Photutils, an Astropy package for detection and photometry of astronomical sources \citep{2020zndo...4049061B}.

%%%%%%%%%%%%%%%%%%%%%%%%%%%%%%%%%%%%%%%%%%%%%%%%%%
\section*{Data Availability}

The Liverpool Telescope {\it RINGO3} data presented here is available through the publicly accessible Liverpool Telescope archive: \url{https://telescope.livjm.ac.uk/cgi-bin/lt_search}



\bibliographystyle{mnras}
%\bibliography{/Users/justyn/Dropbox/main.bib} % if your bibtex file is called example.bib

%\begin{thebibliography}{99}
%\end{thebibliography}

%%%%%%%%%%%%%%%%%%%%%%%%%%%%%%%%%%%%%%%%%%%%%%%%%%

%%%%%%%%%%%%%%%%% APPENDICES %%%%%%%%%%%%%%%%%%%%%

%\appendix

%%%%%%%%%%%%%%%%%%%%%%%%%%%%%%%%%%%%%%%%%%%%%%%%%%

% Don't change these lines
\bsp	% typesetting comment
\label{lastpage}
\end{document}